\documentclass[11pt]{article}

\usepackage{graphicx}
\usepackage{indentfirst}
\usepackage{amssymb}
\usepackage{cite}
\usepackage{color}
\usepackage{subfigure}
\usepackage{varwidth}
\usepackage[colorlinks=true, linkcolor=red, citecolor=blue, urlcolor=magenta]{hyperref}
\usepackage{xcolor}
\usepackage{amsmath}
\usepackage{bm}
\usepackage[normalem]{ulem}

\begin{document}
	
\title{Horizon-scale intensity and polarization images of rotating Konoplya--Zhidenko black holes with thick accretion flows}

\date{}
\maketitle

\begin{center}
\author{Bing-Bing Chen,}$^{a}$\footnote{E-mail: chenbingbing323512@163.com}
\author{Chen-Yu Yang,}$^{b}$\footnote{E-mail: chenyu\_yang2024@163.com}
\author{Guo-Ping Li,}$^{c}$\footnote{E-mail: gpliphys@yeah.net (Corresponding author)}
\author{Xin-Yun Hu}$^{d}$\footnote{E-mail: hu\_xinyun@126.com}
\\

\vskip 0.25in
$^{a}$\it{School of Mathematics, Physics and Statistics, Sichuan Minzu College,   Ganzi 626100, China}\\
$^{b}$\it{Department of Mechanics, Chongqing Jiaotong University, Chongqing 400000, China}\\
$^{c}$\it{Physics and Astronomy College, China West Normal University, Nanchong 637000, China}\\
$^{d}$\it{College of Economics and Management, Chongqing Normal University, Chongqing 401331, China}\\
\end{center}

\vskip 0.6in
{\abstract{
We investigate the shadow and polarization images of a Konoplya--Zhidenko rotating non-Kerr black hole surrounded by a geometrically thick and optically thin accretion flow. The accretion flow is described by an analytical ballistic approximation accretion flow model. The numerical results show that the shadow image exhibits two main features, an outer bright ring and an inner dark region. The former corresponds to higher-order images, while the latter is associated with photon capture by the black hole. Increasing the deformation parameter $\eta$ does not significantly change the overall shape of the higher-order images, but it enlarges their size. Increasing either the spin parameter $a$ or the observer inclination angle $\theta_o$ enhances the asymmetry of the higher-order images and makes the intensity on the left side much larger than that on the right side. This behavior is associated with frame dragging and the relativistic Doppler effect. In the polarization images, the degree of linear polarization is much smaller in the higher-order image region than in other regions, and the polarization vectors extend over the whole image plane. Within the framework used in this work, these results illustrate how the spacetime geometry and near-horizon accretion dynamics jointly influence the intensity and polarization images.
}}

\thispagestyle{empty}
\newpage
\setcounter{page}{1}


\section{Introduction}
A key prediction of general relativity (GR) is the existence of black holes as highly compact astrophysical objects. Identifying black holes and characterizing their physical properties have therefore become central problems in astronomy and astrophysics. In 2016, the LIGO Scientific Collaboration and the Virgo Collaboration announced the detection of a gravitational wave signal produced by the merger of two black holes with masses of approximately $29\,M_\odot$ and $36\,M_\odot$, providing direct observational evidence for such compact objects \cite{abbott2016properties,abbott2016observation}. The Event Horizon Telescope (EHT) subsequently obtained an image of the supermassive black hole at the center of M87* through $1.3\,\mathrm{mm}$ very-long-baseline interferometry observations \cite{event2019first2,akiyama2019first3,akiyama2019first4}. It later released the image of Sagittarius A* at the center of the Milky Way \cite{event2022first1,akiyama2022first3}. These observations provide important data for studying black hole physics in strong gravitational fields and extreme accretion environments.

In EHT images, the outer bright ring and the central low-intensity region are two prominent features. The former is typically associated with radiation from accreting matter around the black hole and strong gravitational lensing. Although the latter is associated with photon capture by the black hole, its observed morphology also depends strongly on the spatial distribution and radiative properties of the emitting plasma.
Early studies already discussed the photon ring and shadow observables of black holes \cite{virbhadra2000schwarzschild,claudel2001geometry,falcke2000viewing}. Since then, methods for calculating black hole shadows have been developed and refined \cite{perlick2022calculating}. Black hole shadows are now widely used in a variety of physical problems. They can be used as standard rulers to quantify and test cosmological phenomena \cite{tsupko2020black,chen2022superradiant} and gravitational theories \cite{zeng2022shadows,mou2022shadows,he2025viewing,yang2026shadow,bambi2019testing,vagnozzi2023horizon}, to constrain black hole parameters \cite{li2014measuring,wei2019intrinsic,hioki2009measurement}, and to explore fundamental issues such as dark matter and quasinormal modes \cite{haroon2019shadow,konoplya2019shadow,hou2018black}. By contrast, the morphology of the bright ring is mainly controlled by the accretion disk model and the emission mechanism. In 1979, Luminet simulated the image of a Schwarzschild black hole with a thin accretion disk by using a semianalytical method \cite{luminet1979image}. That work showed that the shadow image depends on the properties of the accretion flow and on the outer boundary of the central brightness depression. In Ref.~\cite{gralla2019black}, Gralla et al. studied a Schwarzschild black hole surrounded by a geometrically thin and optically thin accretion disk, showing that the bright ring in the black hole image can be decomposed into direct emission, a lensing ring, and a photon ring. The model proposed by Gralla et al. was later extended to other situations involving matter fields and modified gravity, although these studies were still mainly restricted to spherically symmetric black hole backgrounds \cite{zeng2020influence,peng2021influence,he2022influence,li2021observational,zeng2022effects}. Because strong magnetic fields may exist around black holes, the thin disk model was further extended to rotating Kerr--Melvin black holes \cite{hou2022image}. In that work, the accretion disk was assumed to consist of plasma fluid, and the authors explored how the inner shadow and the critical curve could be used to estimate the magnetic field around the black hole. Based on Ref.~\cite{hou2022image}, several studies of black hole images in modified gravity backgrounds have also been carried out \cite{he2025observational,yang2025shadow,li2025shadow,zeng2025kerr}.

Geometrically thin and optically thin accretion disks are widely used in current studies of black hole shadows. However, EHT and related observations indicate that, in strong gravitational fields, the accretion flow near a supermassive black hole may evolve into a geometrically thick and optically thin structure because vertical cooling is suppressed and matter is compressed \cite{akiyama2019first5,event2022first1,ho1999spectral,narayan1994advection}. In this situation, it is necessary to further consider key physical factors such as the electron number density, electron temperature, and magnetic field structure. Most studies of geometrically thick accretion disks adopt a phenomenological radiatively inefficient accretion flow (RIAF) model. The RIAF model has successfully reproduced the overall image features of M87* \cite{akiyama2019first5}, and it agrees well with general relativistic magnetohydrodynamics (GRMHD) simulations \cite{yuan2003nonthermal,broderick2005frequency,broderick2009imaging,saurabh2026semi,jiang2024shadows}. Nevertheless, RIAF models neglect outflows, nonthermal particles, and the full GRMHD process, which limits their use in polarization studies \cite{yuan2014hot,broderick2009estimating,moscibrodzka2009radiative,yang2026distinguishing2}. At the same time, plasma dynamics on event-horizon scales can now be investigated observationally, making the study of near-horizon magnetofluids increasingly important. To address this problem, Hou et al. developed an analytical model of horizon-scale accretion flows within the GRMHD framework, namely the ballistic approximation accretion flow (BAAF) model \cite{hou2024new,zhang2024imaging}. The model assumes that, close to the horizon, the acceleration of the fluid is mainly governed by gravity. By providing explicit expressions for thermodynamic variables and the magnetic field structure, it gives a physically motivated description of the morphology and dynamics of geometrically thick accretion flows in the near-horizon region.

Since possible deviations from the Kerr geometry need not be tied to a known alternative theory, a model-independent parametrized description of black hole spacetimes provides a useful framework for testing the strong-field geometry without specifying a particular theory of gravity. Ref.~\cite{konoplya2016general} developed a general parametrization for stationary and axisymmetric black holes in metric theories of gravity, using a continued-fraction expansion in a compactified radial coordinate for the radial dependence and a Taylor expansion in the cosine of the polar angle for the polar dependence. Ref.~\cite{konoplya2020general} subsequently showed that only a few parameters are sufficient to estimate essential astrophysical observables for spherically symmetric and slowly rotating black holes, as well as for a subclass of arbitrarily rotating black holes belonging to the Carter family. Together, these developments provide an appropriate conceptual basis for treating a small number of deformation parameters as phenomenological probes of the near-horizon geometry \cite{konoplya2016general,konoplya2020general}.

Within this framework, the Konoplya--Zhidenko rotating non-Kerr metric considered in this work should be interpreted as a phenomenological deformation of the Kerr metric characterized by a single deformation parameter $\eta$ \cite{konoplya2016detection}. It reduces to the Kerr metric when $\eta=0$. For nonzero $\eta$, the deformation leaves the asymptotic structure unchanged and therefore preserves the leading weak-field behavior, while modifying the spacetime geometry in the strong-field region near the event horizon \cite{long2018energy}. At the level of the metric ansatz, the Konoplya--Zhidenko metric can be embedded into the broader class of parametrized stationary and axisymmetric black hole geometries because its line element follows from the same general form through a particular choice of the metric functions \cite{konoplya2016general}. For fixed $M$ and $a$, all departures from the Kerr metric are controlled by $\eta$, so varying this parameter spans only a restricted one-dimensional subset of the broader deformation parameter space, with other independent deformations excluded. Accordingly, in this work we use this metric as a phenomenological test geometry and do not associate it with the field equations of any specific alternative theory of gravity.
	
Previous studies have investigated strong gravitational lensing \cite{wang2016strong}, energy extraction through the Penrose process \cite{zeng2026repetitive}, viscous accretion flows \cite{patra2024general}, and numerical methods for charged-particle dynamics \cite{he2023application} in the Konoplya--Zhidenko spacetime. Ref.~\cite{wang2026imaging} studied the intensity and polarization images of spherically symmetric Konoplya--Zhidenko black holes surrounded by geometrically thick accretion flows and found that a central dark region persists in the resulting images. Since that study considered a spherically symmetric spacetime, it did not include the effects of black hole spin. Motivated by this limitation, this paper investigates the shadow and polarization images of a Konoplya--Zhidenko rotating non-Kerr black hole in the BAAF model.

The rest of this paper is organized as follows. Sec.~\ref{sec2} reviews the Konoplya--Zhidenko black hole and the ray-tracing method used for imaging. Sec.~\ref{sec3} introduces the configuration of the BAAF model, including the electron emission model and the accretion disk model. Sec.~\ref{sec4} presents the polarization formalism. Sec.~\ref{sec5} shows the numerical results for the intensity distribution and polarization images. Finally, Sec.~\ref{sec6} gives the conclusions and discussion. Unless otherwise stated, we use geometrized units and adopt the signature convention $(-,+,+,+)$.

\section{Review of the Konoplya--Zhidenko Rotating Non-Kerr Black Hole}\label{sec2}

This section briefly reviews the Konoplya--Zhidenko rotating non-Kerr metric and provides the basic framework for the following analysis. The deformation parameter $\eta$ characterizes a phenomenological deviation from the Kerr metric. In Boyer-Lindquist (BL) coordinates, the metric can be written as \cite{konoplya2016detection}
\begin{align}
	ds^2 &= -\frac{A^2(r,\theta) - B^2(r,\theta) \sin^2\theta}{C^2(r,\theta)} dt^2 
	- 2 B(r,\theta) r \sin^2\theta \, dt \, d\varphi \nonumber\\
	&\quad + C^2(r,\theta) r^2 \sin^2\theta \, d\varphi^2 
	+ \Sigma(r,\theta) r^2 d\theta^2 
	+ \frac{\Sigma(r,\theta) D^2(r,\theta)}{A^2(r,\theta)} dr^2, \label{eq:ma}
\end{align}
where the relevant functions are defined by
\begin{align}
	A^2(r,\theta) &= \frac{r^2 + a^2 - 2 M r}{r^2} - \frac{\eta}{r^3}, \\
	B(r,\theta) &= \frac{2 M a}{r^2 + a^2 \cos^2\theta} 
	+ \frac{a \eta}{r^2 (r^2 + a^2 \cos^2\theta)}, \\
	C^2(r,\theta) &= \frac{(r^2 + a^2)^2 - a^2 \sin^2\theta (r^2 + a^2 - 2 M r)}{r^2 (r^2 + a^2 \cos^2\theta)} 
	+ \frac{a^2 \eta \sin^2\theta}{r^3 (r^2 + a^2 \cos^2\theta)}, \\
	D(r,\theta) &= 1, \quad 
	\Sigma(r,\theta) = \frac{r^2 + a^2 \cos^2\theta}{r^2}.
\end{align}
In this spacetime, $M$ denotes the black hole mass, $a$ is the spin parameter, and $\eta$ is the deformation parameter. When $\eta = 0$, the metric reduces to the Kerr metric. The parameter $\eta$ has only a small effect on the asymptotic structure at infinity, so it is difficult to distinguish a Konoplya--Zhidenko black hole from a Kerr black hole in the weak field regime. The gravitational wave observation GW150914 indicates that, even if a small nonzero deviation exists, the ringdown frequency of a Konoplya--Zhidenko black hole may remain consistent with that of a Kerr black hole \cite{abbott2016observation,abbott2016tests}. However, the presence of $\eta$ modifies the spacetime structure near the event horizon, leading to more pronounced deviations from the Kerr metric in the strong-field region. The event horizon of the metric \eqref{eq:ma} is determined by
\begin{equation}
	\Delta = r^2 - 2 M r + a^2 - \frac{\eta}{r} = 0.
\end{equation}
The value of $\eta$ determines the number and position of the horizons, in clear contrast to the Kerr case. Ref.~\cite{wang2016strong} discussed the influence of the deformation parameter $\eta$ on the event horizon and on the redshift surface at infinity. If the black hole spacetime contains an event horizon, $\eta$ must satisfy \cite{wang2017shadow,wang2024ring}
\begin{equation}
	\begin{cases} 
		\eta \ge \eta_{c} = -\frac{2}{27} \big(\sqrt{4 M^2 - 3 a^2} + 2 M\big)^2 \big(\sqrt{4 M^2 - 3 a^2} - M\big), & a < M, \\
		\eta > 0, & a > M. \label{eq:xi}
	\end{cases}
\end{equation}
For values of $\eta$ outside these ranges, no horizon exists and a naked singularity appears in the spacetime. In accordance with the weak cosmic censorship conjecture, we choose parameters in the following analysis so that an event horizon is present. 
It is worth noting that Eq.~\eqref{eq:xi} shows that, for a Konoplya--Zhidenko black hole, the spin parameter $a$ can be larger than the black hole mass $M$. This represents a significant difference from the Kerr case. The possibility of $a>M$ is a feature of the phenomenological metric adopted in this work rather than a generic prediction of modified gravity. For simplicity, we set $M=1$ in the following.

We next discuss the motion of photons near a Konoplya--Zhidenko black hole. In the spacetime \eqref{eq:ma}, the Hamiltonian of a particle propagating along a geodesic is
\begin{equation}
	H = \frac{1}{2} g^{\mu\nu} p_\mu p_\nu = -\frac{u^2}{2},
\end{equation}
where $p_\mu$ is the particle four-momentum and $u$ is the particle mass. For photons, $u = 0$. Since the Hamiltonian $H$ does not depend on $t$ or $\varphi$, there are two Killing vector fields, $\partial_t$ and $\partial_\varphi$, with the corresponding conserved quantities
\begin{equation}
	E = -p_t = -g_{tt} \dot{t} - g_{t\varphi} \dot{\varphi}, \quad 
	L = p_\varphi = g_{\varphi\varphi} \dot{\varphi} + g_{\varphi t} \dot{t},
\end{equation}
where ``$\dot{}$'' denotes differentiation with respect to the affine parameter $\tau$, and $E$ and $L$ are the energy and angular momentum, respectively. Using these two conserved quantities, the differential equations for photon geodesics can be derived as
\begin{align}
	\dot{t} &= E + \frac{\left(a^2 E-a L+r^2 E\right)\left(2 r^2+\eta\right)}{r^5 \Sigma(r,\theta) A^2(r,\theta)}, \label{eq:ge1}\\
	\dot{\varphi} &= \frac{a E \sin^2\theta\left(2 r^2+\eta\right)+a^2 L r \cos^2\theta-L\left(2 r^2-r^3+\eta\right)}{r^5 \Sigma(r,\theta) A^2(r,\theta)\sin^2\theta}, \\
	\Sigma^2(r,\theta) r^4 \dot{r}^{\,2} &= R(r)=-r^2 A^2(r,\theta)\left[Q+\left(a E-L\right)^2\right]+\left[a L-\left(a^2+r^2\right)E\right]^2, \\
	\Sigma^2(r,\theta) r^4 \dot{\theta}^{\,2} &= \Theta(\theta)=Q+a^2 E^2\cos^2\theta-\frac{L^2\cos^2\theta}{\sin^2\theta}, \label{eq:ge4}
\end{align}
where $Q$ is the generalized Carter constant. The following discussion of black hole shadows relies on Eqs.~\eqref{eq:ge1}--\eqref{eq:ge4}, which accurately describe photon motion near the black hole.

After specifying photon motion, we introduce the ray-tracing method needed to generate black hole shadow images. The detailed procedure can be found in Refs.~\cite{yang2026distinguishing,hu2021qed,zhong2021qed,wang2026semianalytical,he2025shadow,yang2026observational2}. Here we only give a brief summary. A local orthonormal tetrad can be constructed near a zero angular momentum observer (ZAMO) at infinity
\begin{align}
	e_0 &= e_{(t)} = \left(\sqrt{\frac{-g_{\varphi \varphi}}{g_{tt} g_{\varphi\varphi} - g^2_{t \varphi}}}, 0, 0, -\frac{g_{t \varphi}}{g_{\varphi \varphi}} \sqrt{\frac{-g_{\varphi \varphi}}{g_{tt} g_{\varphi\varphi} - g^2_{t \varphi}}} \right), \\
	e_1 &= -e_{(r)} = \left(0, -\frac{1}{\sqrt{g_{rr}}}, 0, 0\right), \\
	e_2 &= e_{(\theta)} = \left(0, 0, \frac{1}{\sqrt{g_{\theta \theta}}}, 0\right), \\
	e_3 &= -e_{(\varphi)} = \left(0, 0, 0, -\frac{1}{\sqrt{g_{\varphi\varphi}}}\right).
\end{align}
In this frame, the photon four-momentum is
\begin{equation}
	p_{(\mu)} = p_{\nu} e_{(\mu)}^{\nu},
\end{equation}
where $p_\nu$ is the four-momentum in BL coordinates. On this basis, celestial coordinates $(\alpha, \beta)$ are introduced. The stereographic projection method gives
\begin{equation}
	\cos\alpha = \frac{p^{(1)}}{p^{(0)}}, \quad \tan\beta = \frac{p^{(3)}}{p^{(2)}}.
\end{equation}
The Cartesian coordinates $(\tilde{x}, \tilde{y})$ on the image screen are related to the celestial coordinates by
\begin{equation}
	\tilde{x} = -2 \tan\frac{\alpha}{2} \sin\beta, \quad
	\tilde{y} = -2 \tan\frac{\alpha}{2} \cos\beta. \label{eq:co1}
\end{equation}
Ray-tracing requires pixel-by-pixel imaging. The image plane is therefore divided into $n \times n$ pixels, and each pixel is labeled by $(i, j)$. The correspondence between Cartesian coordinates and pixel coordinates is
\begin{equation}
	\tilde{x} = l\left(i - \frac{n+1}{2}\right), \quad
	\tilde{y} = l\left(j - \frac{n+1}{2}\right), \label{eq:co2}
\end{equation}
where $l$ is the side length of each pixel. Comparing Eqs.~\eqref{eq:co1} and \eqref{eq:co2}, one obtains the relation between the pixel coordinates and the celestial coordinates
\begin{align}
	\tan\frac{\alpha}{2} &= \frac{1}{n} \tan\left(\frac{\gamma_{\mathrm{fov}}}{2}\right) \sqrt{\left(i - \frac{n+1}{2}\right)^2 + \left(j - \frac{n+1}{2}\right)^2}, \label{eq:c1} \\
	\tan\beta &= \frac{2i - (n+1)}{2j - (n+1)}, \label{eq:c2}
\end{align}
where $\gamma_{\mathrm{fov}}$ is the field of view. {\color{black}The screen coordinates $(\tilde{x},\tilde{y})$ in Eqs.~\eqref{eq:co1}--\eqref{eq:c2} are dimensionless. For plotting, we use the small-angle approximation
\begin{equation}
	(x,y)=
	\frac{r_o\gamma_{\mathrm{fov}}}
	{2\tan(\gamma_{\mathrm{fov}}/2)}(\tilde{x},\tilde{y})
	\simeq r_o(\tilde{x},\tilde{y}).
\end{equation}
The screen coordinates $(x,y)$ are expressed in units of $M$. For a fixed angular field of view, increasing $r_o$ enlarges the region covered by the image.}

\section{Configuration of the Thick Accretion Disk}\label{sec3}

We adopt the BAAF model proposed by Hou et al. \cite{hou2024new,zhang2024imaging}. This model assumes that the fluid acceleration in the near-horizon region is mainly governed by gravity. It also provides analytical expressions for the hot electron temperature, density, and magnetic field configuration, which can be used to study synchrotron imaging at millimeter wavelengths and the effect of anisotropic emission on black hole shadows. This section presents the electron emission model in detail and specifies the physical configuration of the thick accretion disk. The symbols $e$, $c$, $h$, and $k_{B}$ denote the elementary charge, the speed of light in vacuum, the Planck constant, and the Boltzmann constant, respectively.

\subsection{Electron Radiation Model}
In the framework of general relativity, the radiative transfer of unpolarized light obeys the invariant form
\begin{equation}
	\frac{d}{d\tau}\tilde{I} = \tilde{J} - \tilde{\alpha} \tilde{I}, \label{eq:rte}
\end{equation}
where $\tilde{I}, \tilde{J}, \tilde{\alpha}$ are invariant quantities satisfying
\begin{equation}
	\tilde{I} = \frac{I_{\nu}}{\nu^{3}}, \quad \tilde{J} = \frac{j_{\nu}}{\nu^{2}}, \quad \tilde{\alpha} = \nu \alpha_{\nu}. \label{eq:rcs}
\end{equation}
Here $I_{\nu}, j_{\nu}, \alpha_{\nu}$ are the specific intensity, emissivity, and absorption coefficient at the photon frequency $\nu$ in the local reference frame, respectively. In geometrized units, the integral solution of Eq.~(\ref{eq:rte}) is
\begin{equation}
	\tilde{I}(\tau) = \tilde{I}(\tau_0) + \int_{\tau_0}^{\tau} d\tau' \, \tilde{J}(\tau')
	e^{- \int_{\tau'}^{\tau} d\tau'' \, \tilde{\alpha}(\tau'') }.
\end{equation}
To convert geometrized units into CGS units, we introduce the coefficient $k = r_g / \nu_0$ in front of the affine parameter in Eq.~(\ref{eq:rte}), where $r_g = GM / c^2$ is the unit length and $\nu_0$ is the physical photon frequency at infinity. Eq.~(\ref{eq:rte}) then becomes
\begin{equation}
	\frac{1}{k} \frac{d}{d\tau} \tilde{I} = \tilde{J} - \tilde{\alpha} \tilde{I},
\end{equation}
and its solution gives the specific intensity
\begin{equation}
	I_{\nu} = g^{3} I_{\nu_0} + r_g \int_{\tau_0}^{\tau} d\tau' \, g^{2} j_{\nu}(\tau')
	e^{- r_g \int_{\tau'}^{\tau} d\tau'' \, \alpha_{\nu}(\tau'') / g}, \label{eq:is}
\end{equation}
where the redshift factor $g$ is defined as
\begin{equation}
	g = \frac{\nu_0}{\nu} = \frac{k_t}{k_\mu u^\mu} = \frac{-1}{k_\mu u^\mu}.
\end{equation}
In this expression, $k_\mu$ is the photon four-momentum and $u^\mu$ is the fluid four-velocity.

To compute the emissivity and absorption coefficient, we use CGS units in the rest of this section. In the ultrarelativistic regime, the synchrotron emissivity of electrons can be written as
\begin{equation}
	j_{\nu} = \frac{\sqrt{3} e^{3} b \sin{\phi_b}}{4 \pi m_e c^{2}} \int_0^{\infty} d\gamma \, \tilde{N}(\gamma) \tilde{F}\left( \frac{\nu}{\nu_s} \right),\label{eq:j}
\end{equation}
where $b$ is the local magnetic field strength and $\phi_b$ is the angle between the magnetic field orientation and the photon propagation direction. It satisfies
\begin{equation}
	\phi_b = \arccos\left(e_{(b)}^{\mu}e_{(k)\mu}\right).
\end{equation}
In the local reference frame, the normalized vectors along the photon propagation direction and the magnetic field orientation are defined as
\begin{equation}
	e_{(k)}^{\mu} = -\left( \frac{k^{\mu}}{u^\nu k_\nu} + u^\mu \right), \quad e_{(b)}^{\mu} = \frac{b^\mu}{b}.
\end{equation}
In Eq.~\eqref{eq:j}, $\gamma = 1 / \sqrt{1 - \beta^2}$ is the electron Lorentz factor, and the function $\tilde{F}(x)$ is
\begin{equation}
	\tilde{F}(x) = x \int_x^{\infty} dy \, \tilde{K}_{5/3}(y),
\end{equation}
where $\tilde{K}_n(x)$ is the modified Bessel function. The characteristic frequency is
\begin{equation}
	\nu_s = \frac{3 e b \sin\phi_b \, \gamma^2}{4 \pi m_e c}.
\end{equation}
For a thermal electron distribution, the electron distribution function is
\begin{equation}
	\tilde{N}(\gamma) = \frac{n_e \gamma^2 \beta}{\theta_e \tilde{K}_2(1/\theta_e)} e^{-\gamma / \theta_e},
\end{equation}
where $n_e$ is the electron number density, $\theta_e = k_B \tilde{T}_e / m_e c^2$ is the dimensionless electron temperature, and $\tilde{T}_e$ is the electron thermodynamic temperature. In the ultrarelativistic limit ($\theta_e \gg 1$), the modified Bessel function can be approximated as $\tilde{K}_2(1/\theta_e) \approx 2 \theta_e^2$. Introducing the dimensionless variable $q = \gamma / \theta_e$, the emissivity becomes
\begin{equation}
	j_{\nu} = \frac{\sqrt{3} n_e e^3 b \sin{\phi_b}}{8 \pi m_e c^2} \int_0^\infty dq \, q^2 e^{-q} \tilde{F}\left( \frac{\nu}{\nu_s} \right).
\end{equation}
With $x = (\nu / \nu_s) q^2$, this can be further written as
\begin{equation}
	j_{\nu} = \frac{n_e e^2 \nu}{2 \sqrt{3} c \theta_e^2} \tilde{F}(x), \quad x = \frac{\nu}{\nu_c}, \quad \nu_c = \frac{3 e b \sin{\phi_b} \theta_e^2}{4 \pi m_e c}, \label{eq:em}
\end{equation}
where
\begin{equation}
	\tilde{F}(x) = \frac{1}{x} \int_0^\infty dq \, q^2 e^{-q} \tilde{F}\left( \frac{x}{q^2} \right).
\end{equation}
For convenience, this function can be fitted by elementary functions as \cite{mahadevan1996harmony}
\begin{equation}
	\tilde{F}(x) = 2.5651 \left( 1 + 1.92 x^{-1/3} + 0.9977 x^{-2/3} \right) e^{-1.8899 x^{1/3}}. \label{eq:F}
\end{equation}
The model corresponding to Eq.~(\ref{eq:F}) is referred to as anisotropic emission. For the absorption coefficient $\alpha_{\nu}$, the absorption process in a thermal electron distribution obeys Kirchhoff's law, so
\begin{equation}
	\alpha_{\nu} = \frac{j_{\nu}}{\tilde{b}_{\nu}}, \quad
	\tilde{b}_{\nu} = \frac{2 h \nu^3}{c^2} \frac{1}{e^{\,h \nu / (k_B \tilde{T}_e)} - 1},
	\label{eq:kl}
\end{equation}
where $\tilde{b}_{\nu}$ is the Planck blackbody radiation function.

\subsection{Accretion Disk Model}

In the BAAF model, the plasma is assumed to be fully ionized into electrons and protons while remaining electrically neutral. The accreting matter is constrained in the $\theta$ direction by $u^\theta = 0$, and the four-velocity is chosen as a radial accretion flow \cite{pu2016effects,takahashi2011constraining,vincent2022images}
\begin{equation}
	u^\mu = (-g^{tt}, -\sqrt{-(1+g^{tt}) g^{rr}}, 0, -g^{t\varphi}).
\end{equation}
The mass conservation equation is
\begin{equation}
	\frac{d}{dr} \left( \sqrt{-g} \, \rho \, u^r \right) = 0,
\end{equation}
with the solution
\begin{equation}
	\rho = \rho_h \frac{\left. \sqrt{-g} \, u^r \right|_{r=r_h}}{\sqrt{-g} \, u^r},
\end{equation}
where $\rho$ is the mass density, $g$ is the metric determinant, and $r_h$ is the event horizon radius. The projection of the energy-momentum tensor along the four-velocity $u^\mu$ satisfies
\begin{equation}
	d \tilde{U} = \frac{\tilde{U} + p}{\rho} \, d\rho, \label{eq:dedp}
\end{equation}
where $\tilde{U}$ is the internal energy of the fluid. Defining the proton-to-electron temperature ratio as $z = \tilde{T}_p / \tilde{T}_e$, the internal energy of the fluid in this approximation is
\begin{equation}
	\tilde{U} = \rho + \rho \frac{3}{2} (z + 2) \frac{m_e}{m_p} \theta_e, \label{eq:U}
\end{equation}
and the ideal gas equation of state gives
\begin{equation}
	p = n k_B (\tilde{T}_p + \tilde{T}_e) = \rho (1+z) \frac{m_e}{m_p} \theta_e. \label{eq:p}
\end{equation}
Substituting Eqs.~(\ref{eq:U}) and (\ref{eq:p}) into Eq.~(\ref{eq:dedp}) and integrating, we obtain
\begin{equation}
	\theta_e = (\theta_e)_h \left( \frac{\rho}{\rho_h} \right)^{\frac{2(1+z)}{3(2+z)}}.
\end{equation}
The subscript $h$ denotes quantities evaluated at the event horizon $r_h$. In the numerical calculation, $\rho(r_h, \theta)$ is set to a Gaussian distribution in the $\theta$ direction, and $\theta_e(r_h, \theta)$ is taken to be constant in the conical solution
\begin{equation}
	\rho(r_h, \theta) = \rho_h e^{- \left[ (\sin \theta - 1)/0.2 \right]^2 }, \quad \theta_e(r_h, \theta) = \theta_h, \label{eq:th}
\end{equation}
For M87*, observations indicate that $\rho_{h}\approx 1.5\times10^3\,\mathrm{erg\,cm^{-3}}$ and $\theta_h \approx 16.86$ \cite{vincent2022images}. In a stationary and axisymmetric spacetime, the magnetic field is \cite{ruffini1975relativistic}
\begin{equation}
	b^\mu = \frac{\Phi}{\sqrt{-g} \, u^r} \left[ (u_t + \omega_b u_\varphi) u^\mu + \delta_t^\mu + \omega_b \delta_\varphi^\mu \right],
\end{equation}
where
\begin{equation}
	\Phi = \Phi_0 \, \mathrm{sign}(\cos \theta) \, \sin \theta, \quad
	\omega_b = \frac{0.3 a}{2 r_h}.
\end{equation}
Numerical studies show that this magnetic field structure can form naturally in the near-horizon region and can evolve from an initially uniform magnetic field \cite{komissarov2004general,komissarov2004electrodynamics}.

\section{Configuration of Polarization}\label{sec4}

To obtain the image at the observer position, radiation must be propagated from the source to the observer screen. Unlike Eq.~(\ref{eq:rte}), we use the covariant form of the radiative transfer equation to describe the interaction between light rays and matter \cite{gammie2012formalism}
\begin{equation}
	k^\mu \nabla_\mu \tilde{S}^{\alpha\beta} = \tilde{J}^{\alpha\beta} + \tilde{H}^{\alpha\beta\mu\nu} \tilde{S}_{\mu\nu},\label{eq:rte2}
\end{equation}
where $k^\mu$ and $\tilde{S}^{\alpha\beta}$ are the wave vector of the light ray and the polarization tensor, respectively. The tensor $\tilde{J}^{\alpha\beta}$ describes the emission from the radiation source, while $\tilde{H}^{\alpha\beta\mu\nu}$ contains absorption effects and Faraday rotation effects \cite{huang2024coport}. Using the gauge invariance of $\tilde{S}^{\alpha\beta}$, the covariant radiative transfer equation (\ref{eq:rte2}) can be decomposed into two parts. The first part describes gravitational effects
\begin{equation}
	k^\mu \nabla_\mu f^\nu = 0, \quad f^\mu k_\mu = 0,
\end{equation}
where $f^\mu$ is the polarization vector that is parallel transported along the photon trajectory. The second part describes plasma effects
\begin{equation}
	\frac{d}{d\tau} \tilde{S} = \tilde{R}(\tilde{\theta}) \tilde{J} - \tilde{R}(\tilde{\theta}) \tilde{M} \tilde{R}(-\tilde{\theta}) \tilde{S},
\end{equation}
where
\begin{equation}
	\tilde{S}=
	\begin{pmatrix}
		\tilde{I}\\ \tilde{Q}\\ \tilde{U}\\ \tilde{V}
	\end{pmatrix},\quad
	\tilde{J} = \frac{1}{\nu^2}
	\begin{pmatrix}
		j_I\\ j_Q\\ j_U\\ j_V
	\end{pmatrix},\quad
	\tilde{M} = \nu
	\begin{pmatrix}
		\alpha_I & \alpha_Q & \alpha_U & \alpha_V\\
		\alpha_Q & \alpha_I & r_V & -r_U\\
		\alpha_U & -r_V & \alpha_I & r_Q\\
		\alpha_V & r_U & -r_Q & \alpha_I
	\end{pmatrix}.
	\label{eq:ptc}
\end{equation}

To determine the polarized radiative transfer coefficients in Eq.~\eqref{eq:ptc}, we introduce an appropriate local orthonormal tetrad~\cite{broderick2004covariant}
\begin{align}
	\hat{e}^{\mu}_{(0)} &= u^\mu,\quad
	\hat{e}^{\mu}_{(1)} =
	\frac{\epsilon^{\mu\nu\sigma\rho}u_\nu k_\sigma b_\rho}
	{\varpi\mathcal{N}},\\
	\hat{e}^{\mu}_{(2)} &=
	\frac{1}{\mathcal{N}}
	\left(
	b^\mu+\zeta u^\mu-\xi\hat{e}^{\mu}_{(3)}
	\right),\quad
	\hat{e}^{\mu}_{(3)} =
	\frac{k^\mu}{\varpi}-u^\mu,
\end{align}
where
\begin{equation}
	b^2=b_\mu b^\mu,\qquad
	\zeta=u_\mu b^\mu,\qquad
	\varpi=-k_\mu u^\mu,\qquad
	\xi=\frac{k_\mu b^\mu}{\varpi}-\zeta,\qquad
	\mathcal{N}=\sqrt{b^2+\zeta^2-\xi^2}.
\end{equation}
In this tetrad, the polarized radiative transfer coefficients satisfy
$j_U=\alpha_U=r_U=0$. The remaining nonzero emissivity coefficients are given by~\cite{melrose1980plasma}
\begin{align}
	j_I &=
	\frac{n_e e^2\nu}
	{2\sqrt{3}c\theta_e^2}\,
	\tilde{F}(x),\\
	j_Q &=
	\frac{n_e e^2\nu}
	{2\sqrt{3}c\theta_e^2}\,
	\tilde{F}_Q(x),\\
	j_V &=
	\frac{2n_e e^2\nu\cot\phi_b}
	{3\sqrt{3}c\theta_e^3}\,
	\tilde{F}_V(x),
\end{align}
where $\tilde{F}(x)$ is the fitting function given in Eq.~\eqref{eq:F}, and $\tilde{F}_Q(x)$ and $\tilde{F}_V(x)$ are the fitting functions for the linearly and circularly polarized emissivities, respectively, with their explicit forms adopted from Refs.~\cite{dexter2016public,wang2026imaging2}. Under the assumption of local thermodynamic equilibrium, the absorption coefficients are obtained from Kirchhoff's law as
\begin{equation}
	\alpha_I=\frac{j_I}{\tilde{b}_{\nu}},\qquad
	\alpha_Q=\frac{j_Q}{\tilde{b}_{\nu}},\qquad
	\alpha_V=\frac{j_V}{\tilde{b}_{\nu}}.
\end{equation}
The Faraday conversion coefficient $r_Q$ and Faraday rotation coefficient $r_V$ are given by~\cite{shcherbakov2008propagation,dexter2016public}
\begin{align}
	r_Q &=
	\frac{n_e e^4 b^2\sin^2\phi_b}
	{4\pi^2m_e^3c^3\nu^3}
	\mathcal{F}_m(X)
	\left[
	\frac{\tilde{K}_1(1/\theta_e)}
	{\tilde{K}_2(1/\theta_e)}
	+6\theta_e
	\right],\\
	r_V &=
	\frac{n_e e^3 b\cos\phi_b}
	{\pi m_e^2c^2\nu^2}
	\frac{
		\tilde{K}_0(1/\theta_e)-\mathcal{J}_5(X)
	}{
		\tilde{K}_2(1/\theta_e)
	},
\end{align}
where
\begin{equation}
	X=
	\left(
	\frac{3}{2\sqrt{2}}\times10^{-3}\frac{\nu}{\nu_c}
	\right)^{-1/2},
\end{equation}
and $\mathcal{F}_m(X)$ and $\mathcal{J}_5(X)$ are the fitting functions for the Faraday coefficients, whose explicit forms are adopted from Ref.~\cite{dexter2016public}.

The rotation angle $\tilde{\theta}$ is defined as the angle between $f^\mu$ and $b^\mu$
\begin{equation}
	\tilde{\theta} = \mathrm{sign}(\epsilon_{\alpha\beta\mu\nu} u^\alpha f^\beta b^\mu k^\nu)
	\arccos \left( \frac{h^{\mu\nu} f_\mu b_\nu}{\sqrt{(h^{\mu\nu} f_\mu f_\nu)(h^{\alpha\beta} b_\alpha b_\beta)}} \right),
\end{equation}
where $h^{\mu\nu}$ is the induced metric on the transverse subspace. The corresponding rotation matrix is
\begin{equation}
	\tilde{R}(\tilde{\theta}) =
	\begin{pmatrix}
		1 & 0 & 0 & 0 \\
		0 & \cos (2\tilde{\theta}) & -\sin (2\tilde{\theta}) & 0 \\
		0 & \sin (2\tilde{\theta}) & \cos (2\tilde{\theta}) & 0 \\
		0 & 0 & 0 & 1
	\end{pmatrix}.
\end{equation}
At the observer, the Stokes parameters are projected onto the observer’s image plane. The corresponding rotation angle is \cite{zhou2026nonthermal}
\begin{equation}
	\tilde{\theta}_o = 
	\mathrm{sign}(\epsilon_{\alpha\beta\mu\nu} u^{\alpha} f^{\beta} d^{\mu} k^{\nu})
	\arccos\left(
	\frac{h^{\mu\nu} f_\mu d_\nu}{\sqrt{(h^{\mu\nu} f_\mu f_\nu)(h^{\alpha\beta} d_\alpha d_\beta)}}
	\right),
\end{equation}
where $d^{\mu}=-\partial_\theta^\mu$. The projected Stokes parameters are
\begin{equation}
	\tilde{I}_o = \tilde{I},\quad
	\tilde{Q}_o = \tilde{Q}\cos(2\tilde{\theta}_o) - \tilde{U}\sin(2\tilde{\theta}_o),\quad
	\tilde{U}_o = \tilde{Q}\sin(2\tilde{\theta}_o) + \tilde{U}\cos(2\tilde{\theta}_o),\quad
	\tilde{V}_o = \tilde{V}.
\end{equation}
The Stokes parameters determine the magnitude and orientation of the linear polarization vector $\vec{f}$ on the observer’s image plane. Its magnitude corresponds to the degree of linear polarization, while its orientation is characterized by the electric vector position angle (EVPA)
\begin{equation}
	P_o = \frac{\sqrt{\tilde{Q}_o^{\,2} + \tilde{U}_o^{\,2}}}{\tilde{I}_o}, \qquad
	\Theta_o = \frac{1}{2} \arctan\!\left(\frac{\tilde{U}_o}{\tilde{Q}_o}\right).
	\label{eq:pv}
\end{equation}
The direction and magnitude of the linear polarization vector provide an intuitive description of the spatial distribution of polarization.

\section{Numerical Simulation Results}\label{sec5}

Before presenting the numerical results, we briefly summarize the numerical setup used throughout this section. During ray tracing, the integration is terminated when the ray reaches the near-horizon boundary $r=1.01r_h$, propagates outward beyond the radial cutoff $r=600M$, or {\color{black}the accumulated optical depth reaches $\mathcal{T}=50$, where $\mathcal{T}$ is defined as
\begin{equation}
	\mathcal{T}(\tau)
	=
	r_g \int_0^\tau
	\frac{\alpha_\nu(\tau')}{g(\tau')}\,d\tau',
\end{equation}
with $\tau=0$ at the observer.} We also set the maximum affine parameter to $\tau_{\max}=3000$ as the upper limit of the numerical integration. No additional plasma cutoff is imposed beyond the above termination conditions. {\color{black}The numerical integration uses adaptive step sizes. We checked convergence by increasing the image plane resolution and reducing the maximum allowed integration step size while keeping all physical parameters fixed. The intensity and polarization structures remain stable under further refinement, including near the higher-order images.} For visualization, we apply the following nonlinear mapping to the radiation intensity
\begin{equation}
I_{\rm plot}
=
-\left[
\log_2\left(
\frac{I}{I_{\max}+I}
\right)
\right]^{-1},
\end{equation}
where $I_{\max}$ denotes the maximum radiation intensity in the corresponding image. {\color{black}Each image is normalized independently. Brightness comparisons between images therefore refer to normalized contrast or left-right asymmetry.} The remaining common numerical parameters are listed in Table~\ref{tab1}.

\begin{table}[htbp]
	\centering
	\caption{Common numerical parameters.}
	\label{tab1}
	\begin{tabular}{ccc}
		\hline
		Quantity & Symbol & Value \\
		\hline
		Observing frequency & $\nu_0$ & $230\,\mathrm{GHz}$ \\
		Proton-to-electron temperature ratio & $z=\tilde{T}_p/\tilde{T}_e$ & $20$ \\
		{\color{black}Magnetic parameter} & {\color{black}$\Phi_0$} & {\color{black}$1$ (dimensionless)} \\
		Observer radius & $r_o$ & $500M$ \\
		{\color{black}Image resolution} & {\color{black}$n\times n$} & {\color{black}$256\times256$} \\
		Boundary Stokes parameters & $\tilde{S}_b$ & $(0,0,0,0)^{\mathrm{T}}$ \\
		\hline
	\end{tabular}
\end{table}

{\color{black}The parameter $\Phi_0$ in Table~\ref{tab1} is dimensionless. The physical magnetic field strength used in the calculations is expressed in gauss ($\mathrm{G}$) and is set by
\begin{equation}
	b_{\mathrm{phys}}(r,\theta)
	=
	b_h\sqrt{
		\frac{b^\mu b_\mu(r,\theta)}
		{\left.b^\mu b_\mu\right|_*}
	},
	\quad b_h=10\,\mathrm{G},
\end{equation}
where $b^\mu$ is the magnetic field configuration computed with $\Phi_0=1$, and the subscript $*$ denotes evaluation at the reference position $r_*=1.001r_h$, $\theta_*=\pi/2-0.001$. Thus, $\Phi_0$ cancels in the ratio, and the physical magnetic field is fixed by the field strength $b_h$ at the reference position.}

\subsection{Intensity Images}

\begin{figure}[!h]
	\centering 
	\subfigure[$\eta=-0.9,\theta_o=0^\circ$]{\includegraphics[scale=0.45]{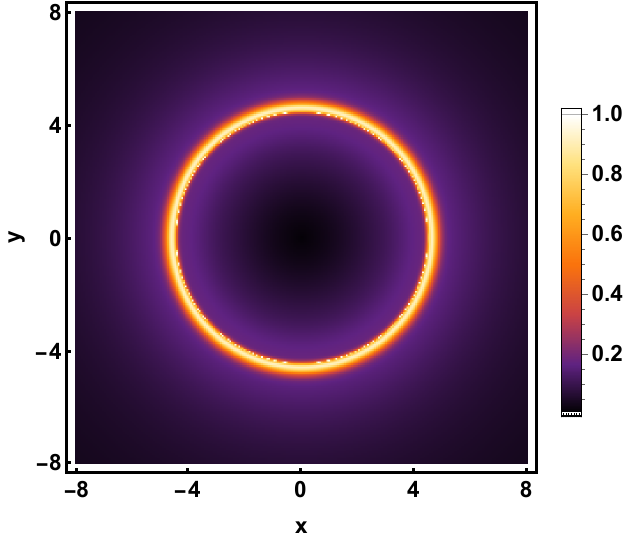}}
	\subfigure[$\eta=-0.9,\theta_o=17^\circ$]{\includegraphics[scale=0.45]{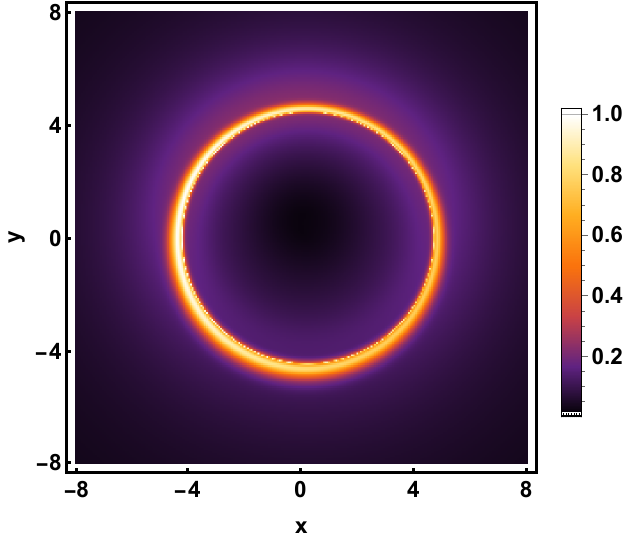}}
	\subfigure[$\eta=-0.9,\theta_o=75^\circ$]{\includegraphics[scale=0.45]{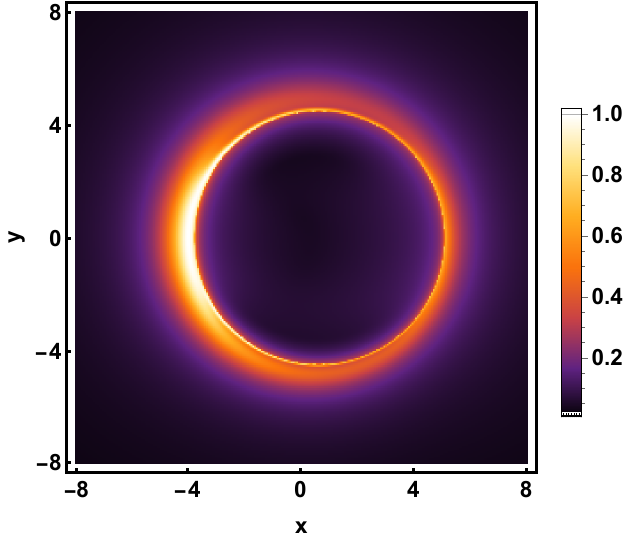}}
	
	\subfigure[$\eta=0.1,\theta_o=0^\circ$]{\includegraphics[scale=0.45]{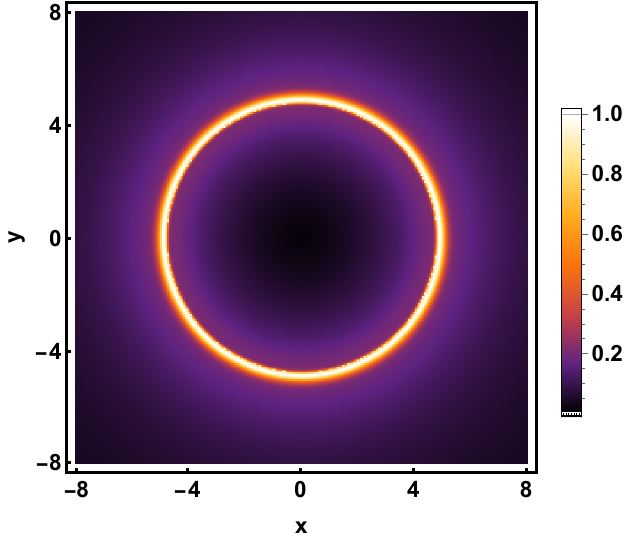}}
	\subfigure[$\eta=0.1,\theta_o=17^\circ$]{\includegraphics[scale=0.45]{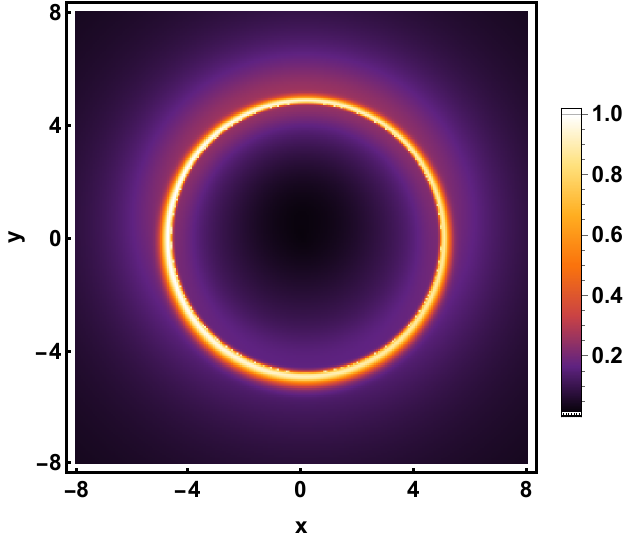}}
	\subfigure[$\eta=0.1,\theta_o=75^\circ$]{\includegraphics[scale=0.45]{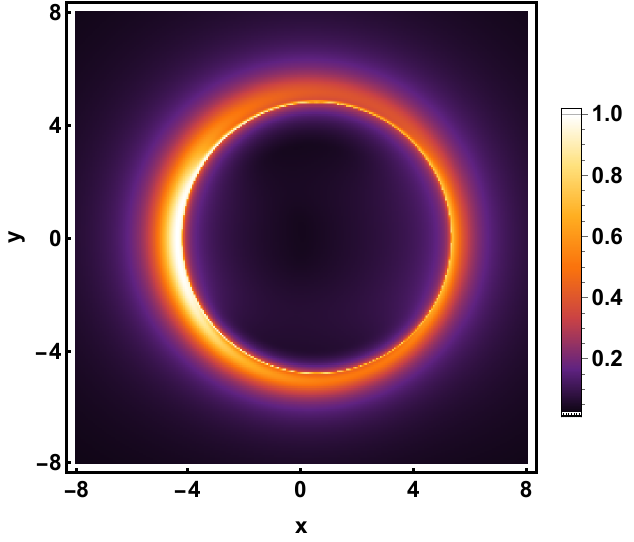}}
	
	\subfigure[$\eta=0.9,\theta_o=0^\circ$]{\includegraphics[scale=0.45]{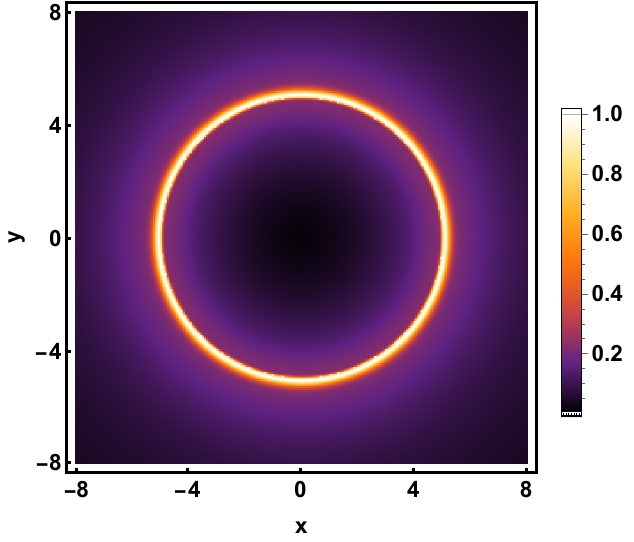}}
	\subfigure[$\eta=0.9,\theta_o=17^\circ$]{\includegraphics[scale=0.45]{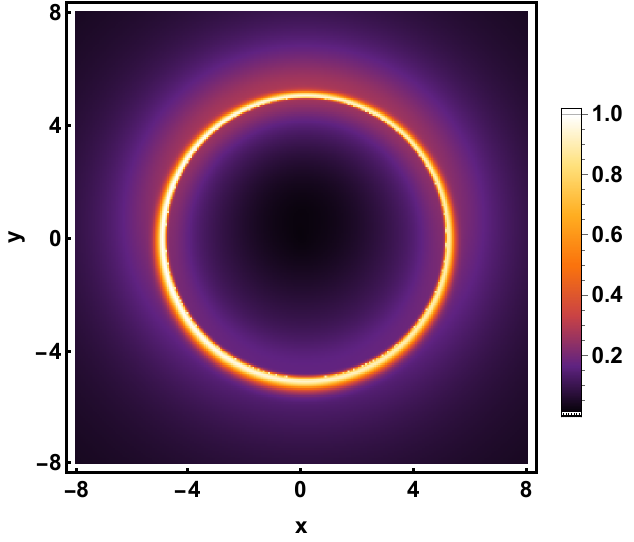}}
	\subfigure[$\eta=0.9,\theta_o=75^\circ$]{\includegraphics[scale=0.45]{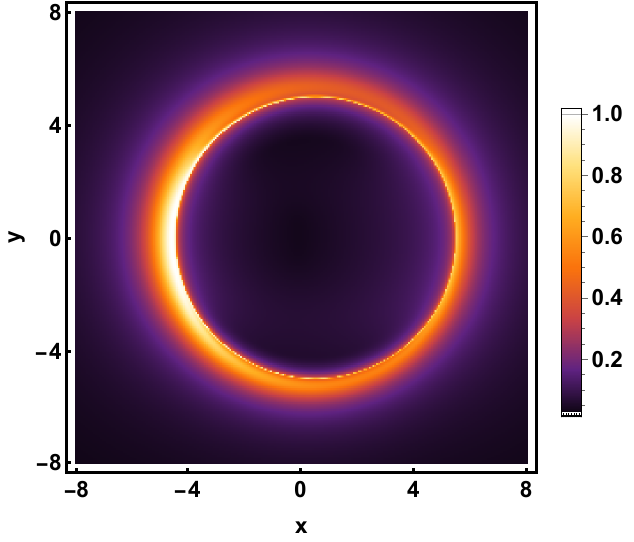}}
	
	\caption{Effects of the deformation parameter $\eta$ and the observer inclination angle $\theta_o$ on the black hole shadow in the BAAF model. The spin parameter is fixed at $a=0.3$.}
	\label{fig1}
\end{figure}

Fig.~\ref{fig1} shows the effects of the deformation parameter $\eta$ and the observer inclination angle $\theta_o$ on the black hole shadow. A central low-intensity region can be seen in all the images. We refer to this radiative feature as the inner dark region, which is associated with photon capture by the black hole. For a geometrically thin accretion disk, the accreting matter lies on the equatorial plane and extends inward to the event horizon. In this case, the lensed image of the equatorial event horizon can appear as a black region with a clear boundary, commonly referred to as the ``inner shadow'', and may be captured by the EHT \cite{chael2021observing}. For a geometrically thick accretion disk, however, emission from outside the equatorial plane can obscure this region, making its boundary blurred and difficult to observe. Another prominent feature is the bright ring-like structure in the image. This structure corresponds to higher-order images, produced by photons that orbit the black hole one or more times before reaching the observer. The higher-order images are a direct manifestation of strong gravitational lensing. Outside the ring-like structure, there remains a region of finite intensity corresponding to the primary image, formed by light rays that travel directly from the accretion disk to the observer.

\begin{figure}[!h]
	\centering 
	\subfigure[$x$ direction]{\includegraphics[scale=0.8]{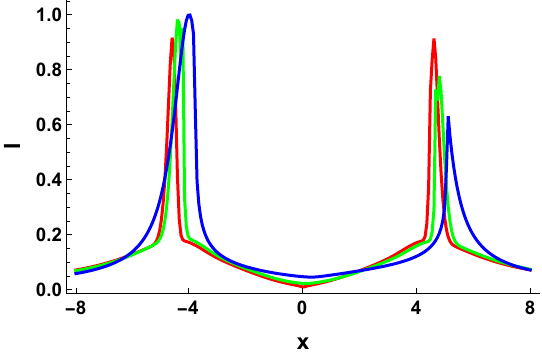}}
	\subfigure[$y$ direction]{\includegraphics[scale=0.8]{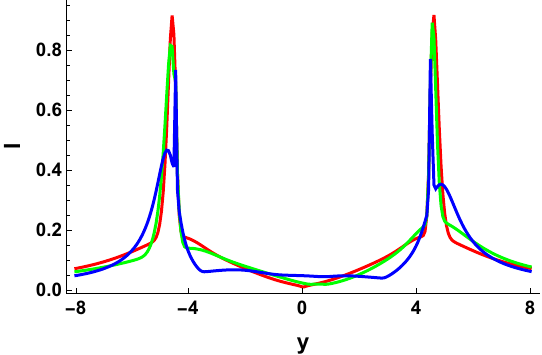}}

	\caption{{\color{black}Normalized intensity cuts along the $x$ and $y$ directions.} The fixed parameters are $\eta=-0.9$ and $a=0.3$, and the red, green, and blue curves correspond to $\theta_o=0^\circ, 17^\circ, 75^\circ$, respectively.}
	\label{fig2}
\end{figure}

\begin{figure}[!h]
	\centering 
	\subfigure[$x$ direction]{\includegraphics[scale=0.8]{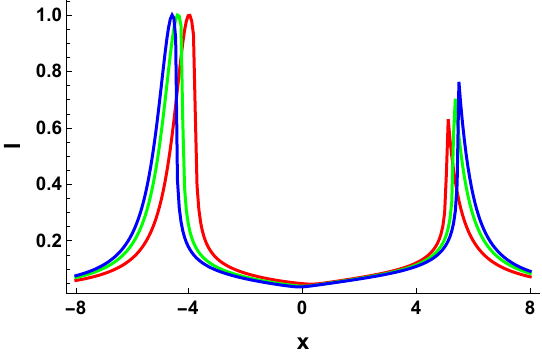}}
	\subfigure[$y$ direction]{\includegraphics[scale=0.8]{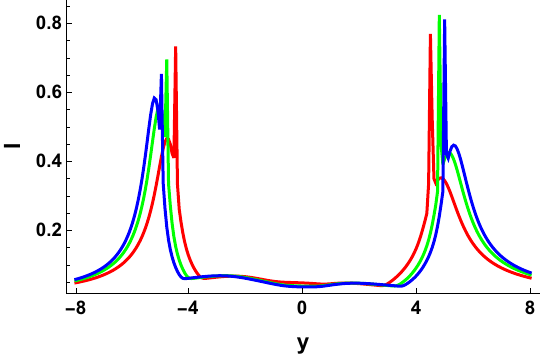}}
	
	\caption{{\color{black}Normalized intensity cuts along the $x$ and $y$ directions.} The fixed parameters are $a=0.3$ and $\theta_o=75^\circ$, and the red, green, and blue curves correspond to $\eta=-0.9, 0.1, 0.9$, respectively.}
	\label{fig3}
\end{figure}

In Fig.~\ref{fig1}, when $\theta_o = 0^\circ$, the higher-order images form an almost perfect circular ring. As $\theta_o$ increases to $17^\circ$, the ring-like structure becomes slightly distorted. When $\theta_o = 75^\circ$, the Doppler effect caused by the relative motion makes the intensity on the left side of the higher-order images much larger than that on the right side. Increasing the deformation parameter $\eta$ has little effect on the overall shape of the higher-order images, but it enlarges their size. {\color{black}To illustrate the normalized intensity distributions more clearly, Figs.~\ref{fig2} and \ref{fig3} show the normalized intensity cuts along the $x$ and $y$ directions. The two pronounced peaks correspond to the higher-order images, while the regions outside the peaks correspond to the primary image. In Fig.~\ref{fig2}, as $\theta_o$ increases, the peak positions along the $x$ direction shift to the right, while those along the $y$ direction remain nearly unchanged. In Fig.~\ref{fig3}, as $\eta$ increases, the two peaks gradually move away from each other, indicating that the size of the higher-order images increases.}

\begin{figure}[!h]
	\centering 
	\subfigure[$a=0.1$]{\includegraphics[scale=0.45]{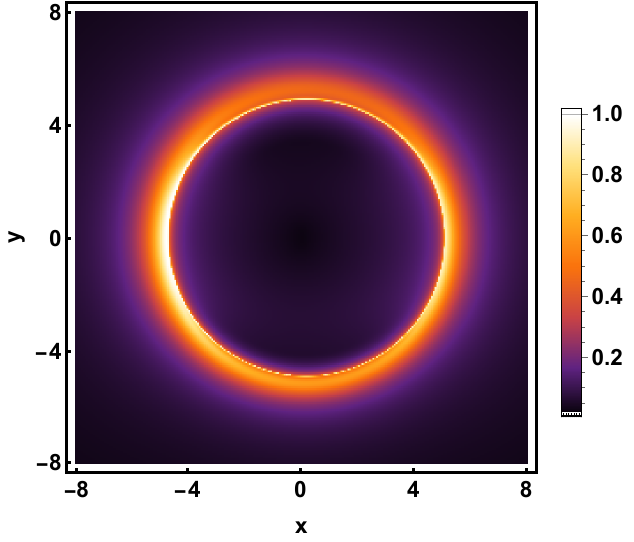}}
	\subfigure[$a=0.7$]{\includegraphics[scale=0.45]{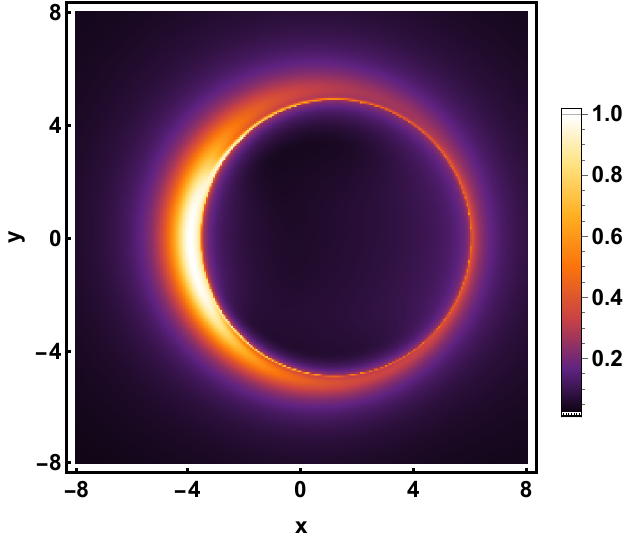}}
	\subfigure[$a=1.2$]{\includegraphics[scale=0.45]{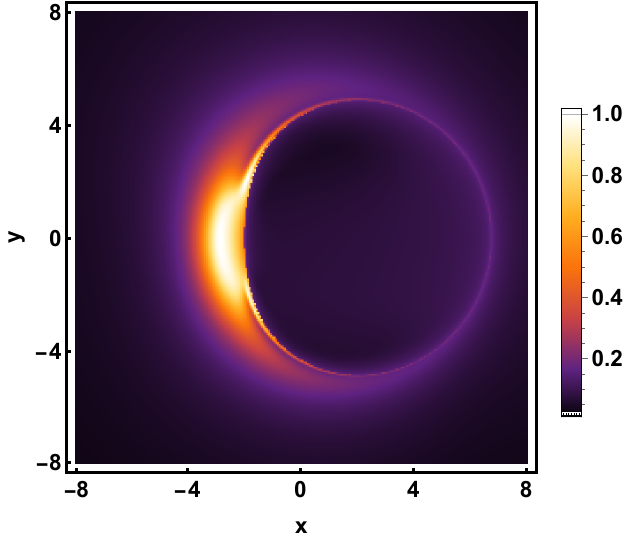}}
	
	\caption{Effect of the spin parameter $a$ on the black hole shadow in the BAAF model. The deformation parameter and observer inclination angle are fixed at $\eta=0.5$ and $\theta_o=75^\circ$, respectively.}
	\label{fig4}
\end{figure}

Fig.~\ref{fig4} shows the effect of the spin parameter $a$ on the black hole shadow image. Unlike a Kerr black hole, a Konoplya--Zhidenko black hole can have a spin parameter $a$ larger than the black hole mass $M=1$. We therefore choose $a=0.1, 0.7, 1.2$. {\color{black}The images show that the left-right brightness asymmetry of the higher-order images increases with $a$, with the left side brighter than the right.} When $a=1.2$, a crescent-shaped bright region appears on the left side, and the higher-order images become D-shaped. This feature originates from the frame-dragging effect. A larger spin parameter $a$ strengthens frame dragging and therefore increases the asymmetry of the higher-order images.

\begin{figure}[!h]
	\centering 
	\subfigure[$a=0.1$]{\includegraphics[scale=0.45]{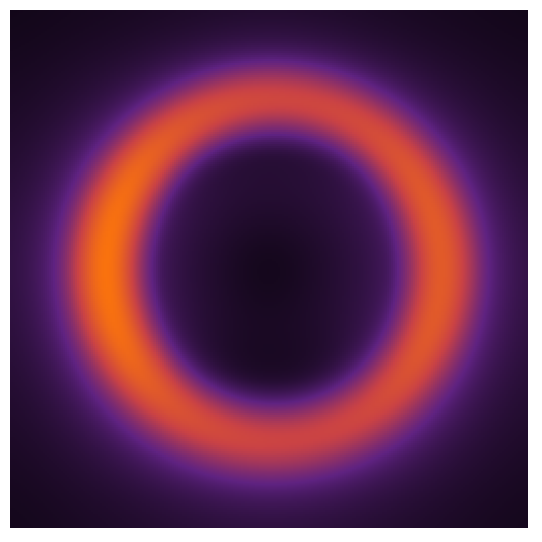}}
	\hspace{0.5cm}
	\subfigure[$a=0.7$]{\includegraphics[scale=0.45]{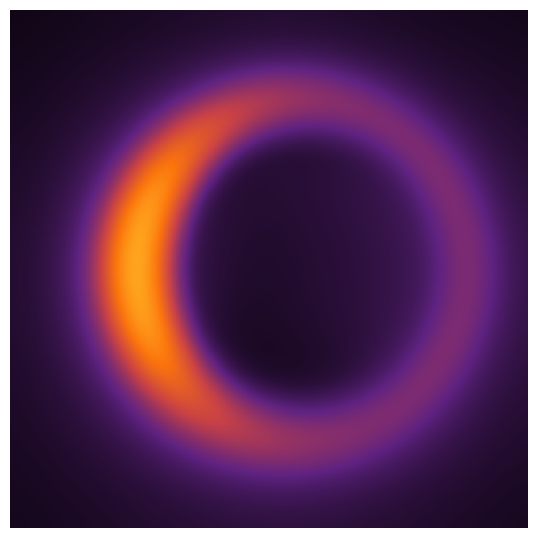}}
	\hspace{0.5cm}
	\subfigure[$a=1.2$]{\includegraphics[scale=0.45]{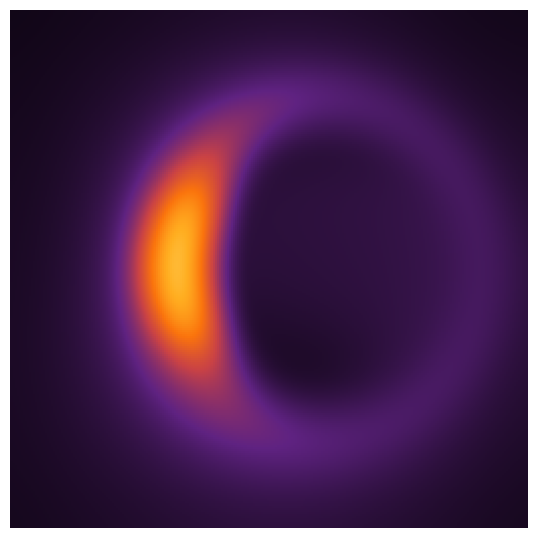}}
	
	\caption{Blurred images processed with a Gaussian filter corresponding to a circular Gaussian beam having an FWHM of $20\,\mu\mathrm{as}$. The remaining parameters are the same as in Fig.~\ref{fig4}.}
	\label{fig5}
\end{figure}

To compare with the limited angular resolution of current EHT observations, Fig.~\ref{fig5} shows the blurred images corresponding to Fig.~\ref{fig4}. We adopt M87* as the reference source, with a black hole mass of $M_{\mathrm{BH}}=6.5\times10^{9}\,M_{\odot}$ and a source distance of $D_{\mathrm{BH}}=16.9\,\mathrm{Mpc}$ \cite{event2019first4}. In the ray-tracing calculation, we set the observer radius to $r_o=500M$ and the numerical camera field of view to $\gamma_{\mathrm{fov}}=2^\circ$. With the adopted mass and source distance of M87*, this numerical setup corresponds to a sky angular field of view of approximately $66.3\,\mu\mathrm{as}\times66.3\,\mu\mathrm{as}$. The intensity maps are then convolved with a circular Gaussian beam having an FWHM of $20\,\mu\mathrm{as}$ \cite{gralla2019black}. In Fig.~\ref{fig5}, the boundary of the inner dark region becomes less distinguishable, and the boundary between the higher-order images and the primary image becomes blurred. Nevertheless, the overall trend of the image variation caused by $a$ is still preserved.

We briefly compare our results with previous studies. Ref.~\cite{wang2017shadow} investigated the geometric shadow of the same rotating Konoplya--Zhidenko spacetime and found that increasing the deformation parameter enlarges the shadow and makes its boundary more rounded. Although the geometric shadow and the higher-order images are distinct observables, their sizes show the same trend with increasing deformation parameter. For the non-rotating Konoplya--Zhidenko spacetime, Ref.~\cite{wang2026imaging} likewise found that the size of the higher-order images increases with the deformation parameter. Since that study considered a non-rotating spacetime, it did not capture the difference in intensity between the left and right sides arising from rotation. Ref.~\cite{yang2026shadow} performed a similar radiative-transfer imaging analysis for rotating black holes in Kalb--Ramond gravity. The authors found that increasing the spontaneous Lorentz-violating parameters reduces the size of the higher-order images, whereas increasing the spin parameter enhances their brightness asymmetry. The latter trend is qualitatively consistent with our result for the effect of rotation.

\subsection{Polarization Images}

\begin{figure}[!h]
	\centering 
	\subfigure[$\tilde{Q}_{o}$]{\includegraphics[scale=0.45]{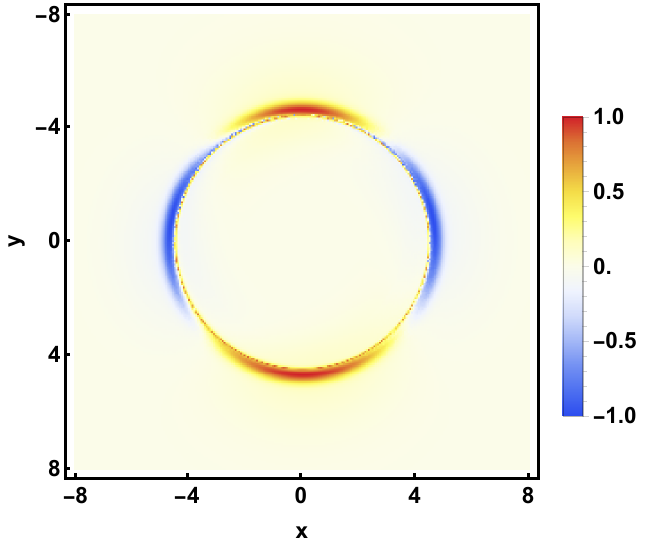}}
	\subfigure[$\tilde{U}_{o}$]{\includegraphics[scale=0.45]{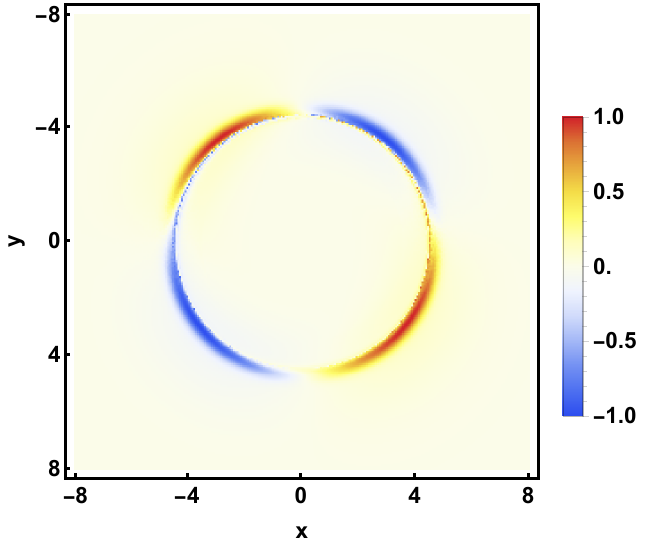}}
	\subfigure[$\tilde{V}_{o}$]{\includegraphics[scale=0.45]{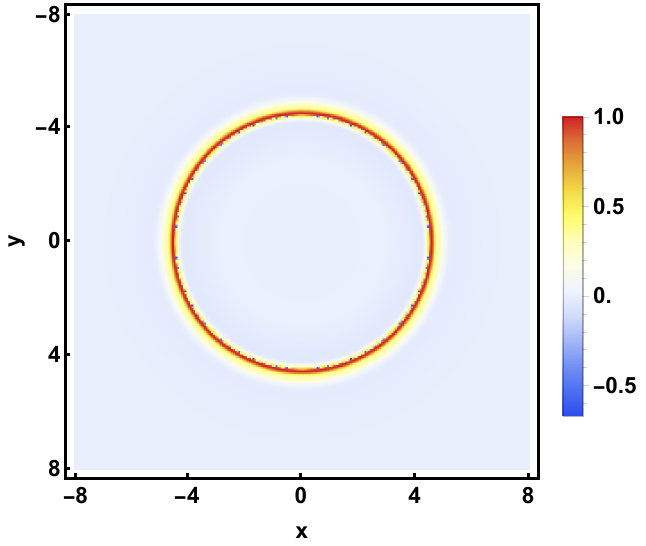}}
	
	\caption{Distributions of the Stokes parameters $\tilde{Q}_{o}, \tilde{U}_{o}, \tilde{V}_{o}$. The fixed parameters are $\eta=-0.9, a=0.3, \theta_o=0^\circ$.}
	\label{fig6}
\end{figure}

This subsection analyzes the polarization images of the Konoplya--Zhidenko black hole in the BAAF model. We study the distributions of the Stokes parameters and the linear polarization vector $\vec{f}$ on the image screen. Fig.~\ref{fig6} shows the spatial distributions of the Stokes parameters $\tilde{Q}_{o}, \tilde{U}_{o}, \tilde{V}_{o}$ for the fixed parameters $\eta=-0.9, a=0.3, \theta_o=0^\circ$. The distributions of $\tilde{Q}_{o}$ and $\tilde{U}_{o}$ are mainly concentrated near the higher-order images. For $\tilde{V}_{o}$, positive values appear near the higher-order images, indicating left-handed circular polarization. The remaining regions have negative values, corresponding to right-handed circular polarization.

\begin{figure}[!h]
	\centering 
	\subfigure[$\eta=-0.9,\theta_o=0^\circ$]{\includegraphics[scale=0.45]{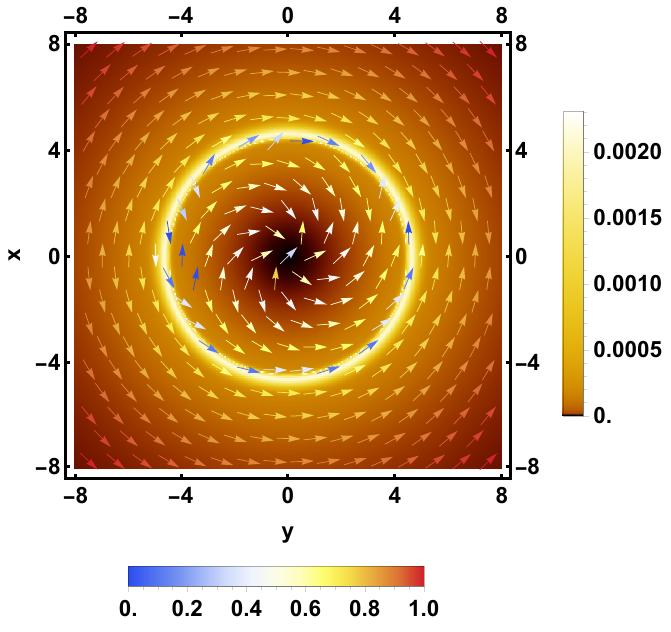}}
	\subfigure[$\eta=-0.9,\theta_o=17^\circ$]{\includegraphics[scale=0.45]{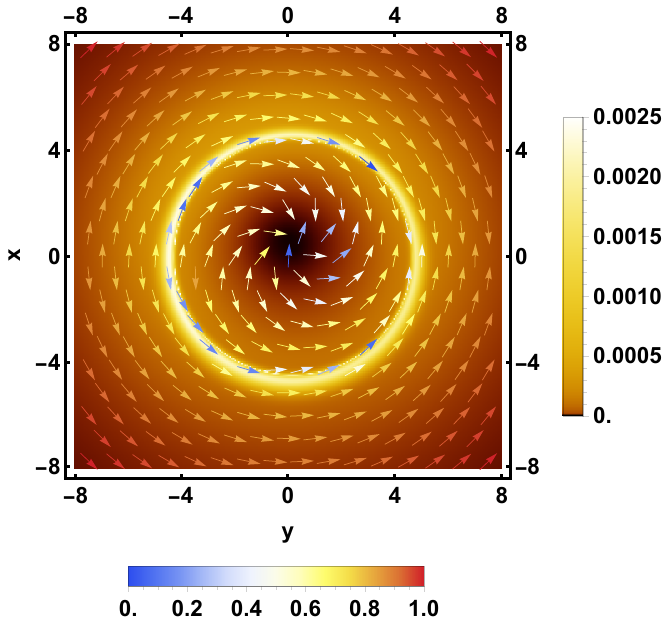}}
	\subfigure[$\eta=-0.9,\theta_o=75^\circ$]{\includegraphics[scale=0.45]{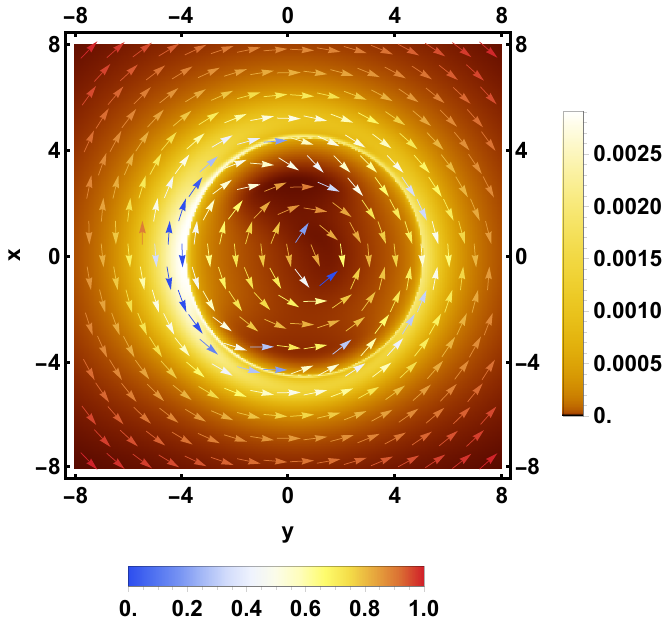}}
	
	\subfigure[$\eta=0.1,\theta_o=0^\circ$]{\includegraphics[scale=0.45]{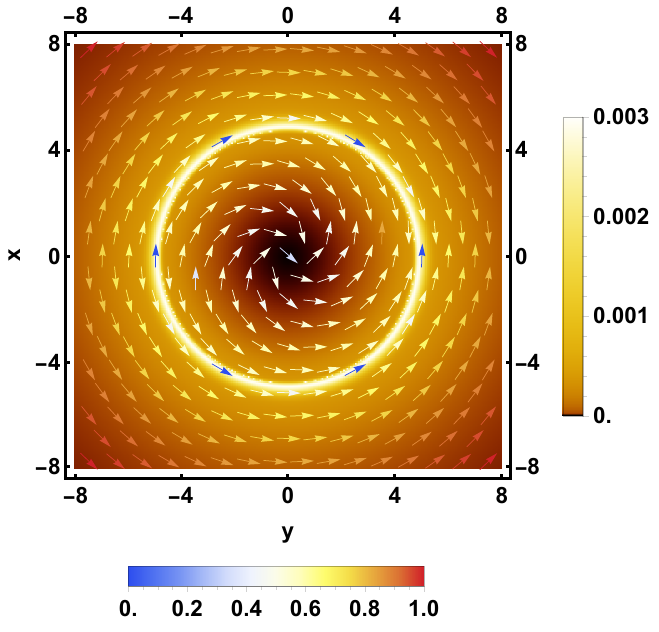}}
	\subfigure[$\eta=0.1,\theta_o=17^\circ$]{\includegraphics[scale=0.45]{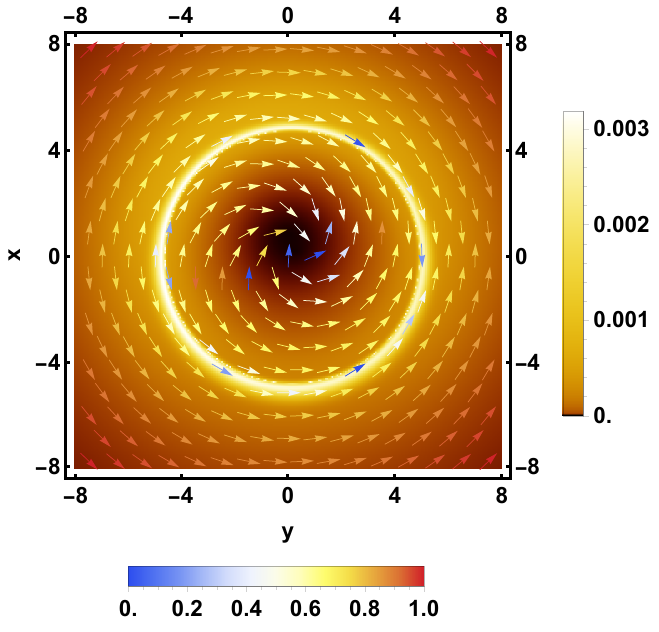}}
	\subfigure[$\eta=0.1,\theta_o=75^\circ$]{\includegraphics[scale=0.45]{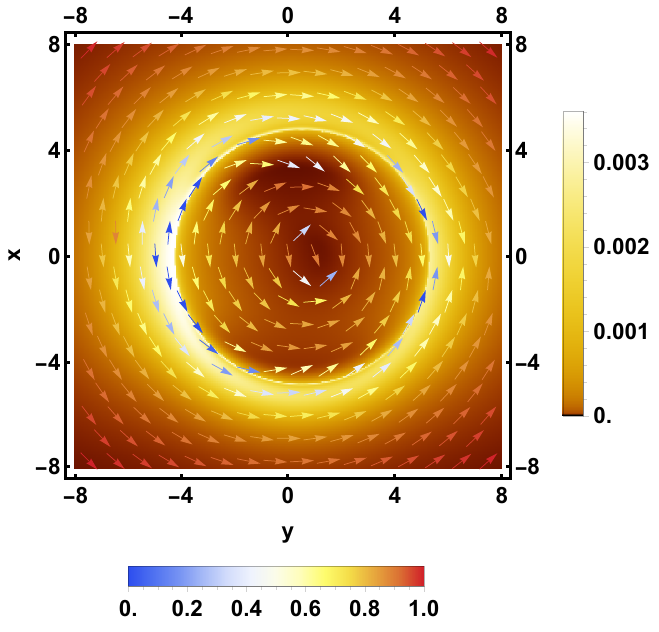}}
	
	\subfigure[$\eta=0.9,\theta_o=0^\circ$]{\includegraphics[scale=0.45]{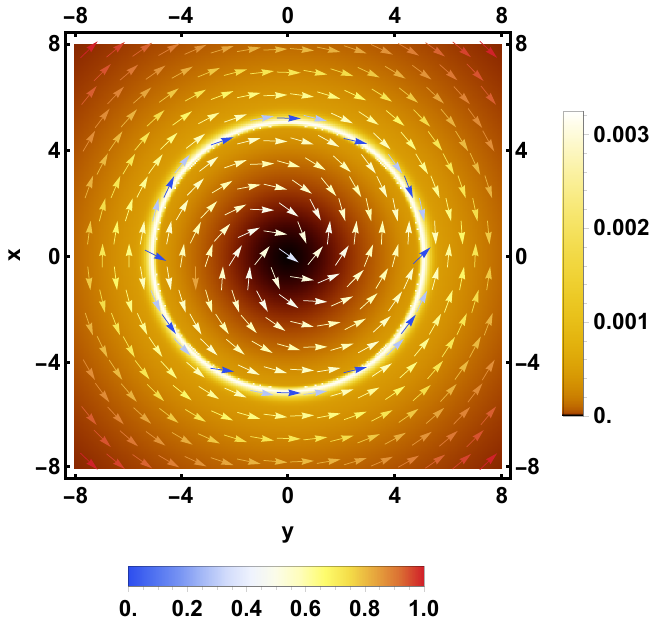}}
	\subfigure[$\eta=0.9,\theta_o=17^\circ$]{\includegraphics[scale=0.45]{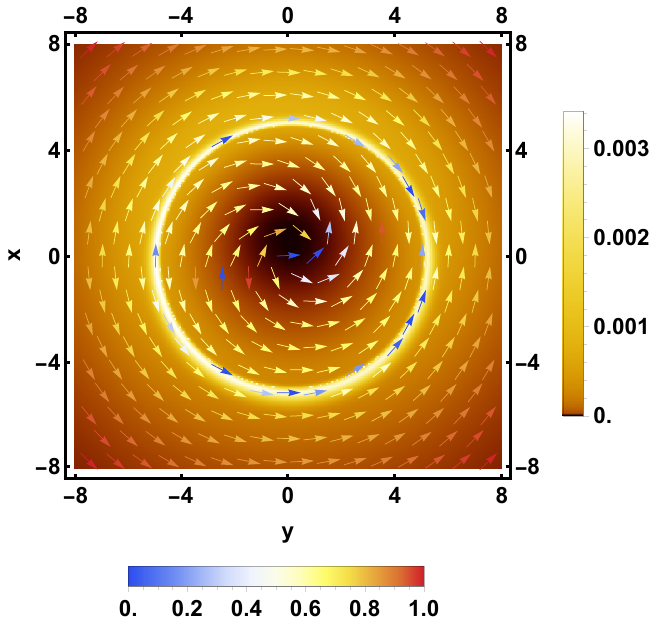}}
	\subfigure[$\eta=0.9,\theta_o=75^\circ$]{\includegraphics[scale=0.45]{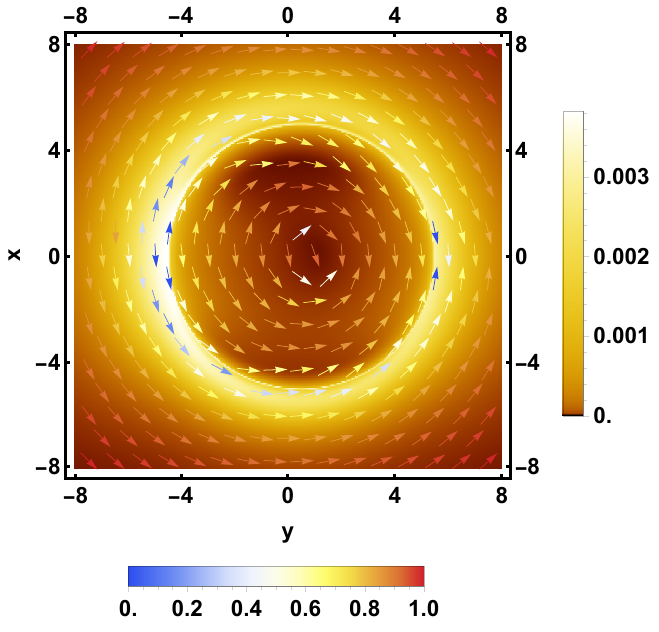}}
	
	\caption{Effects of the deformation parameter $\eta$ and the observer inclination angle $\theta_o$ on the polarization images. The spin parameter is fixed at $a=0.3$.}
	\label{fig7}
\end{figure}

\begin{figure}[!h]
	\centering 
	\subfigure[$a=0.1$]{\includegraphics[scale=0.45]{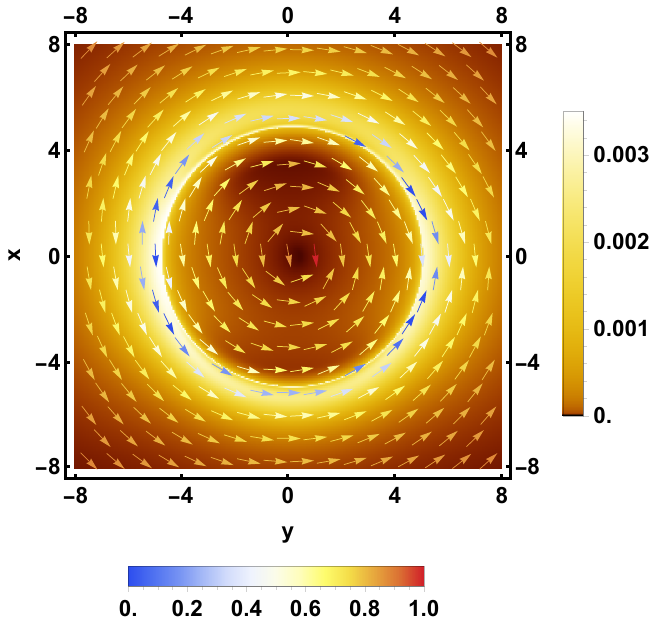}}
	\subfigure[$a=0.7$]{\includegraphics[scale=0.45]{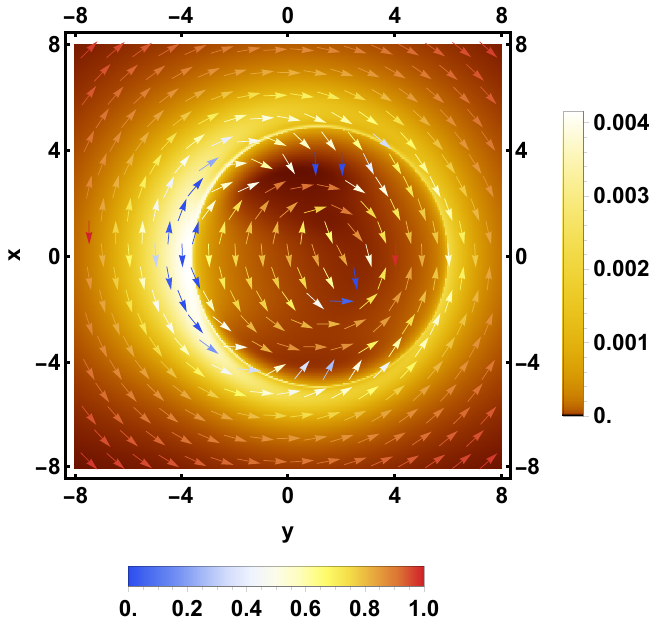}}
	\subfigure[$a=1.2$]{\includegraphics[scale=0.45]{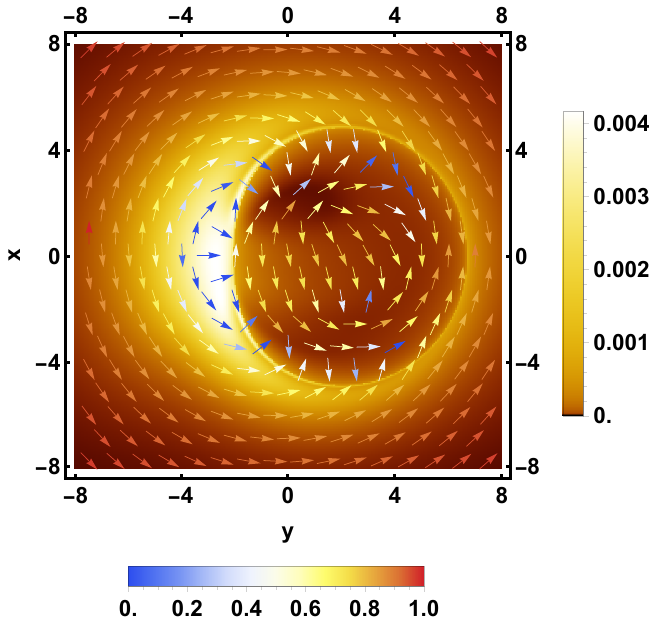}}
	
	\caption{Effect of the spin parameter $a$ on the polarization images. The deformation parameter and observer inclination angle are fixed at $\eta=0.5$ and $\theta_o=75^\circ$, respectively.}
	\label{fig8}
\end{figure}

Figs.~\ref{fig7} and \ref{fig8} show the effects of the black hole parameters $\eta, \theta_o$, and $a$ on the Stokes parameter $\tilde{I}_{o}$ and the linear polarization vector $\vec{f}$. In these figures, $\tilde{I}_{o}$ denotes the intensity distribution. The arrows represent the linear polarization vector, and their colors and directions correspond to the degree of linear polarization $P_{o}$ and the EVPA $\Theta_{o}$, respectively. The images show that $\tilde{I}_{o}$ increases significantly near the higher-order images. According to Eq.~(\ref{eq:pv}), this increase reduces $P_{o}$. Away from the higher-order image region, the intensity decreases and $P_{o}$ increases. Since the EVPA is perpendicular to the magnetic field $\vec{b}$, the polarization pattern can reflect the magnetic field geometry. Far from the black hole, the magnetic field is approximately radial. For Fig.~\ref{fig7}, when the observer inclination angle is small, the distributions of $\vec{f}$ inside and outside the higher-order images differ noticeably. For Fig.~\ref{fig8}, when $a=1.2$, the distribution of $\vec{f}$ becomes disordered near the crescent-shaped bright region on the left side of the higher-order images. It is worth noting that, in the thin disk model, no polarization effect can be observed in the inner shadow region \cite{yang2026observational}. In contrast, for the geometrically thick accretion disk considered here, gravitational lensing allows emission above and below the equatorial plane to partially obscure the inner dark region, producing polarization vectors across the whole image plane.

{\color{black}
For the polarization images, our results are qualitatively consistent with those obtained in Ref.~\cite{yang2026shadow} within the BAAF model. Both studies find that the signals in the Stokes parameters $\tilde{Q}_o$ and $\tilde{U}_o$ are mainly concentrated near the higher-order images. The degree of linear polarization is lower in this region and gradually increases farther away. The polarization analysis in Ref.~\cite{yang2026shadow} fixed the observer inclination angle and spin parameter, focusing mainly on the effects of the KR parameters. We further analyze the effects of $\theta_o$ and $a$ on the polarization vector distribution. As shown in Fig.~\ref{fig7}, the polarization vector distributions inside and outside the higher-order images exhibit discernible differences at small observer inclination angles. Fig.~\ref{fig8} further shows that the EVPA distribution inside the higher-order images changes with increasing spin.
}

\section{Conclusion and Discussion}\label{sec6}

In recent years, many studies have focused on simulations of black hole shadow images illuminated by thin accretion disks. In more realistic physical scenarios, however, the geometrical thickness of the accretion disk should also be taken into account. Motivated by this issue, we studied the shadow images of a Konoplya--Zhidenko rotating non-Kerr black hole illuminated by a thick disk. We first reviewed the basic properties of the Konoplya--Zhidenko black hole, whose deviation from a Kerr black hole is characterized by the deformation parameter $\eta$. We then briefly introduced the differential equations for photon geodesics and the ray-tracing method. An analytical BAAF model was adopted to describe the geometrically thick accretion flow. By numerically integrating the null geodesic equations and the radiative transfer equations, we computed the synchrotron emission from thermal electrons in the magnetofluid and obtained the corresponding shadow and polarization images.

For the shadow images, we analyzed the effects of the deformation parameter $\eta$, the observer inclination angle $\theta_o$, and the spin parameter $a$. The results show that the shadow image is mainly composed of an outer bright ring and an inner dark region. The outer bright ring corresponds to higher-order images and is a direct manifestation of gravitational lensing. The inner dark region is associated with photon capture by the black hole and may be partially obscured by emission away from the equatorial plane. {\color{black}As $\theta_o$ increases, the Doppler effect enhances the left-right brightness asymmetry of the higher-order images, with the left side brighter than the right.} Increasing $\eta$ does not significantly change the overall shape of the higher-order images, but it enlarges their size. These results are clearly reflected in the intensity cuts along the $x$ and $y$ directions. Unlike a Kerr black hole, a Konoplya--Zhidenko black hole can have a spin parameter $a$ larger than the black hole mass because of the deformation parameter $\eta$. We find that increasing $a$ produces a significant frame-dragging effect, deforms the higher-order images into a ``D'' shape, and enhances the image asymmetry. In the blurred images, the higher-order images and the primary image become difficult to distinguish, indicating that direct imaging of structures near the black hole event horizon remains challenging.

In the polarization images, we find that the Stokes parameters $\tilde{Q}_{o}$, $\tilde{U}_{o}$, and $\tilde{V}_{o}$ are mainly concentrated near the higher-order images. The degree of linear polarization $P_o$ is mainly controlled by the Stokes parameter $\tilde{I}_{o}$, and the two are negatively correlated. As a result, $P_o$ is smaller at the higher-order images. The distribution of the EVPA $\Theta_o$ depends strongly on $\theta_o$ and $a$. Smaller $\theta_o$ and larger $a$ lead to different distributions of $\Theta_o$ inside and outside the higher-order images. It is worth noting that, in the thin-disk model, no polarization signal appears in the inner shadow. In contrast, for the geometrically thick accretion disk considered here, emission away from the equatorial plane contributes along these lines of sight, allowing the polarization vectors to extend across the whole image plane. Overall, this study highlights the influence of the Konoplya--Zhidenko black hole parameters on shadow and polarization features within a thick accretion disk model. Since the image morphology is jointly affected by the spacetime geometry, accretion dynamics, and magnetic field configuration, the observational interpretation of the trends found here should be treated with caution. In future work, we will further include the full GRMHD process to simulate more realistic accretion dynamics and radiation processes.


\cleardoublepage

\vspace{10pt}
\noindent {\bf Acknowledgments}

\noindent
This work is supported by the National Natural Science Foundation of China (Grants Nos. 12375043 and 12575069), the Science and Technology Research Program of Chongqing Municipal Education Commission (Grant No. KJZD-K202600503), and the Chongqing Normal University Fund Project (Grant No. 26XLB001).

\bibliographystyle{utphys} 
\bibliography{biblio} 

\end{document}